\documentclass[prl,aps,epsfig,superscriptaddress]{revtex4}
\usepackage{bm}
\usepackage{amsfonts}
\usepackage[dvips]{graphicx}
\usepackage{mathrsfs}
\usepackage[intlimits]{amsmath}
\usepackage[colorlinks, citecolor=red]{hyperref}

\begin{document}

\title{Experimental realization of a multiplexed quantum memory with $225$
individually accessible memory cells}
\author{Y.-F. Pu$^{1}$, N. Jiang$^{1}$, W. Chang$^{1}$, H.-X. Yang$^{1}$, C.
Li$^{1}$, L.-M. Duan}
\affiliation{Center for Quantum Information, IIIS, Tsinghua University, Beijing 100084,
PR China}
\affiliation{Department of Physics, University of Michigan, Ann Arbor, Michigan 48109, USA}

\begin{abstract}
To realize long-distance quantum communication and quantum network, it is
required to have multiplexed quantum memory with many memory cells. Each
memory cell needs to be individually addressable and independently
accessible. Here we report an experiment that realizes a multiplexed
DLCZ-type quantum memory with $225$ individually accessible memory cells in
a macroscopic atomic ensemble. As a key element for quantum repeaters, we
demonstrate that entanglement with flying optical qubits can be stored into
any neighboring memory cells and read out after a programmable time with
high fidelity. Experimental realization of a multiplexed quantum memory with
many individually accessible memory cells and programmable control of its
addressing and readout makes an important step for its application in
quantum information technology.
\end{abstract}

\maketitle

\section{Introduction}

For realization of long-distance quantum communication through the quantum
repeater network \cite{1,2,3,4,5,6} or linear optics quantum computation
using atomic ensembles \cite{7,9}, a key requirement is to have multiplexed
quantum memory with many memory cells. Each memory cell needs to be
individually accessible with access time independent of physical location of
the cells. A well-known implementation of the quantum repeater network is
through the Duan-Lukin-Cirac-Zoller (DLCZ) scheme \cite{2}, where
probabilistically generated atom-photon entanglement is stored in the
DLCZ-type memory \cite{2,3,4,5}. On-demand readout of the information or
entanglement stored in the DLCZ memory at a programable time is critical for
archiving efficient scaling of the quantum repeater scheme \cite{1,2,3,5}.

Atomic internal states in the ground-state manifold provide an ideal
candidate for realization of quantum memory because of their long coherence
time and the possibility of an efficient interface with the quantum bus
carried by the flying photonic qubits \textbf{\cite{3,4}}. In particular,
the collective internal states of a many-atom ensemble have enhanced
coupling to directional light and thus provides a quantum memory with
efficient interface to photons even in free space \textbf{\cite{2,3,4,5}}.
Such an ensemble-based quantum memory has found many applications \textbf{%
\cite{3,4,5}}, in particular for realization of quantum internet and
long-distance quantum communication \textbf{\cite{1,2,3,4,5,6}}. Quantum
memory with efficient interface to light has been demonstrated up to a few
qubits using atomic ensembles in either vacuum or low-temperature crystals
\textbf{\cite{11,12,13,14,15,16,17,18,19,20,21,22}}. Through access to the
orbital angular momentum or spatial imaging space of the photon \textbf{\cite%
{32,33}}, high-dimensional photon-photon entanglement can be generated and
up to seven-dimensional entanglement has been stored in the atomic ensemble
\textbf{\cite{34}}. To increase the memory capacity, a more efficient method
is to use the multiplexed memory of multiple qubits as an $n$-qubit memory
has an effective Hilbert space dimension of $2^{n}$. One can use
multiplexing in the time-bin \textbf{\cite{23,24,31}}, the spatial profile
\textbf{\cite{17}}, or the angular directions \textbf{\cite{30}} to increase
the memory capacity. Multiple time-bin qubits have been mapped into and out
of a solid-state atomic ensemble at pre-determined time \textbf{\cite{23,24}}%
. The read-out for this system, however, is not on demand and we have only
sequential instead of programmable random access to the stored information.
For the multiplexing based on spatial profile or the angular directions,
either the memory capacity is still limited \textbf{\cite{17}} or one has
yet to achieve full quantum-limited operation \textbf{\cite{30}}. To scale
up the capability of quantum memory, which is critical for its application
in quantum information technology \textbf{\cite{3,4,5}}, we need extendable
quantum systems with many memory cells and programmable access to each cell
with location-independent access time similar to the classical random access
memory.

In this paper, we demonstrate a multiplexed quantum memory with $225$
individually accessible memory cells. This is achieved by dividing a
macroscopic atomic ensemble into a two-dimensional (2D) array of
micro-ensembles, each serving as a quantum memory cell. We use crossed
acoustic optical deflectors (AODs) to realize a 2D multiplexing and
demultiplexing optical circuit so that each memory cell can be individually
addressed by the write and read laser beams and the quantum signals from
each cell can be coupled into single mode fibers in a programmable way. AODs
provide a convenient device for design of multiplexing circuits, which have
been used recently for individual addressing of single atoms \textbf{\cite%
{25}}, ions \textbf{\cite{26}}, and one-dimensional (1D) array of ensembles
\textbf{\cite{17}}. The development from 1D array \textbf{\cite{17}} to the
2D geometry of multiplexing is critical for the dramatic increase of the
memory capacity. To verify functionality of our multiplexed quantum memory,
we perform write and read operations on a $15\times 15$ 2D array of memory
cells, and demonstrate their quantum correlations and individual
controllability with negligible crosstalk. In the DLCZ scheme, we typically
use a pair of neighboring memory cells to store entanglement with flying
optical qubits \cite{2,3}. Through quantum state tomography, we demonstrate
that entanglement with flying optical qubits can be stored into any chosen
pair of neighboring memory cells and read out after a programmable time with
about $90\%$ fidelity. This constitutes an important step for realization of
quantum repeaters.

\begin{figure}[tbp]
\includegraphics[width=15cm]{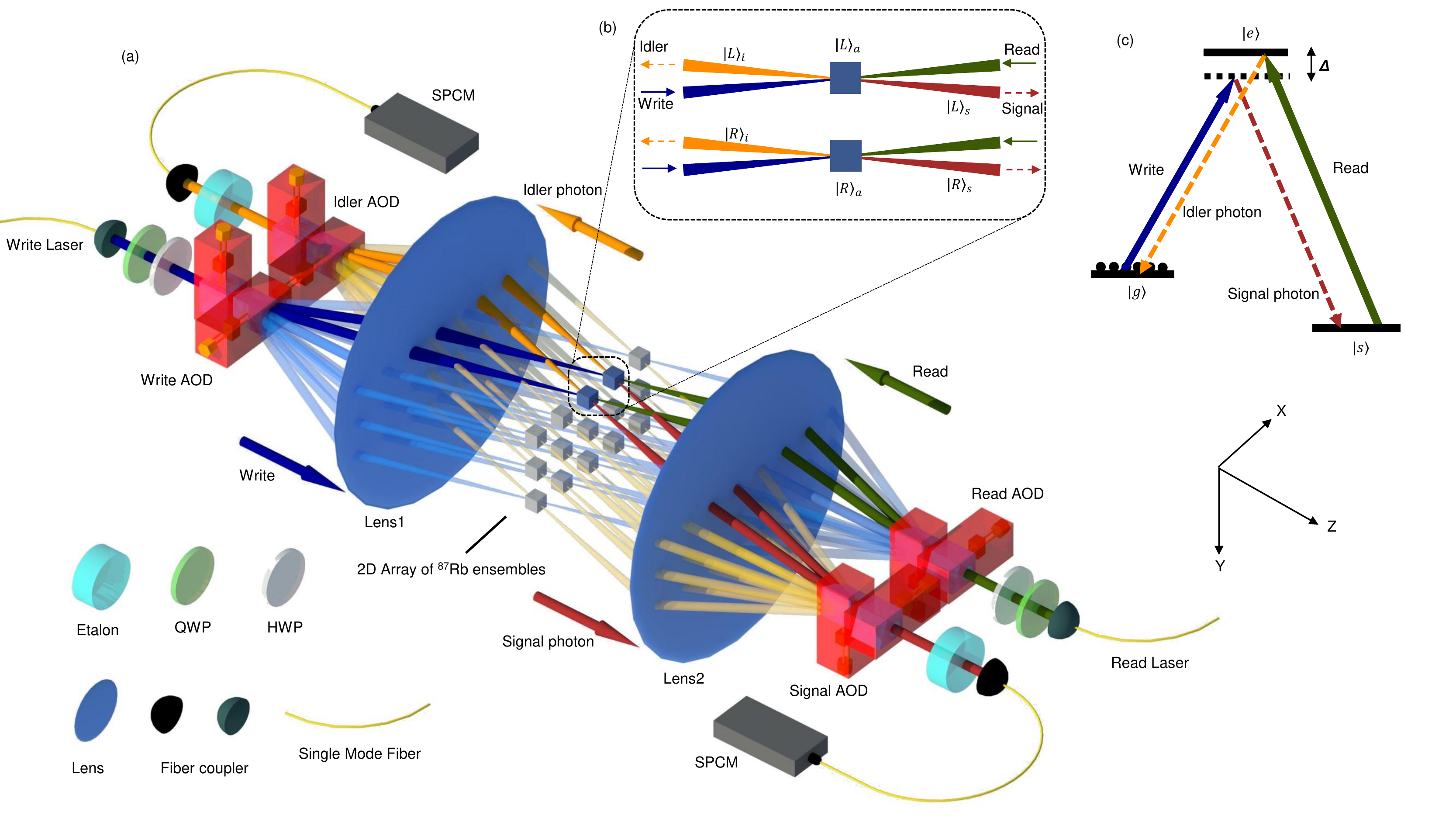}
\caption{\textbf{Experimental setup for demonstration of a multiplexed
quantum memory with $225$ memory cells.} \textbf{a}, The write and the read laser beams are directed by the
multiplexing AODs to address $15\times 15$ atomic memory cells contained in
single macroscopic atomic ensemble, with the distance between neighboring
cells of $126$ $\mu m$. For clarity, only $3\times 5$ cells are shown in the
figure. Write-in/read-out of quantum information to/from the memory cell is
achieved through the DLCZ scheme. The signal and the idler photons emitted
by the memory cells from different paths are combined by the de-multiplexing
AODs into the same single-mode fiber for detection with a single-photon
counting module (SPCM). The polarization of the write/read beam from a
single mode fiber is adjusted by a quarter-wave plate (QWP) and a half-wave
plate (HWP) to generate the highest first-order diffraction efficiency from
the AOD. For each beam, a pair of crossed AODs in orthogonal directions $X$
and $Y$ are used to scan the angle of deflected beam in the corresponding
directions. The lens are used to map different angles of the deflected beams
to different positions at the atomic ensemble as well as to focus the beams.
A Fabry-Perot cavity (etalon) is inserted in the path of the signal/idler
mode to further filter our the strong write/read beam by frequency
selection. \textbf{b}, Zoom-in of the beam configuration at two memory cells
denoted as L and R. \textbf{c}, The relevant level diagram of $^{87}$Rb
atoms for the write and read process, with $\left\vert g\right\rangle \equiv
\left\vert 5S_{1/2},F=2\right\rangle $, $\left\vert s\right\rangle \equiv
\left\vert 5S_{1/2},F=1\right\rangle $, and $\left\vert e\right\rangle
\equiv \left\vert 5P_{1/2},F^{\prime }=2\right\rangle $. The detuning of the
write beam is at $\Delta =10$ MHz.}
\end{figure}

\section{Results}

\subsection{Experimental configuration}

We use the two hyperfine states $\left\vert g\right\rangle \equiv \left\vert
5S_{1/2},F=2\right\rangle $ and $\left\vert s\right\rangle \equiv \left\vert
5S_{1/2},F=1\right\rangle $ of $^{87}$Rb atoms to carry quantum information
for the memory. Our experimental setup is shown in Fig. 1, with details
described in the Methods. All the atoms are initially prepared to the state $%
\left\vert g\right\rangle $ through a repumping laser pulse applied to the
transition $\left\vert s\right\rangle \rightarrow \left\vert
5P_{3/2},F^{\prime }=2\right\rangle $. We use the DLCZ scheme to generate
quantum correlation between a collective mode of the atomic ensemble and the
signal photon \textbf{\cite{2}}, which are propagating in the forward
direction with an angle $2^{\text{o}}$ to the write laser beam. The
excitation stored in the collective mode is retrieved after a controllable
delay time by a read laser beam counter-propagating with the write beam, and
the retrieved idler photon, counter propagating with the signal photon, is
coupled into a single-mode fiber for detection.

To generate many memory cells in a single macroscopic atomic ensemble and
fetch quantum signals from these cells with site-independent access time, we
split the write and the read beams into $15\times 15$ different paths by the
2D multiplexing optical circuit shown in Fig. 1. The corresponding signal
and idler photons in each path are collected by the demultiplexing circuit
on the other side. As explained in the Methods, the multiplexing and
demultiplexing optical circuits are composed by crossed AODs and lens and
can be programed by applying radio-frequency electric signals to direct the
light to any paths or their superpositions. The phase differences between
all these optical paths are intrinsically stable as they go through the same
optical apparatus. We therefore do not require active phase locking between $%
225$ different optical paths, which significantly simplifies the
experimental setup.

\subsection{Characterization of quantum correlation of $225$ memory cells}

To characterize quantum property of each memory cell, first we show the
intensity cross-correlation $g_{\text{c}}$ between the signal and the idler
photons generated from each memory cell. This correlation, defined as $g_{%
\text{c}}=\left\langle E_{\text{s}}^{\dagger }E_{\text{s}}E_{\text{i}%
}^{\dagger }E_{\text{i}}\right\rangle /\left[ \left\langle E_{\text{s}%
}^{\dagger }E_{\text{s}}\right\rangle \left\langle E_{\text{i}}^{\dagger }E_{%
\text{i}}\right\rangle \right] $, gives a signature of nonclassical property
of the signal and the idler optical fields with the field operators $E_{%
\text{s}}$ and $E_{\text{i}}$ when $g_{\text{c}}>2$ \textbf{\cite{27a}}. As
shown in Fig. 2a, this criterion for quantum correlation is clearly
satisfied for all the $225$ memory cells. When we move from the center to
the edge of the ensemble, the cross correction $g_{\text{c}}$ gradually
decreases, which is mainly caused by the reduced optical depth from the
center to the edge. The minimum $g_{\text{c}}$ at the edge is still well
above $10$ and can be further increased if we reduce the excitation
probability of the atomic mode by the write beam.

An important feature of the multi-qubit quantum memory is that different
memory cells can be controlled independently without mutual influence to
each other. The major crosstalk comes form the neighboring cells by the
spread of the write or read laser beams. In Figs. 2(b) and 2(c), we show the
crosstalk errors on the neighboring sites induced by the write and the read
beams. For both cases, the maximum error rate is well below $1\%$.

\begin{figure}[ptb]
\includegraphics[width=17cm]{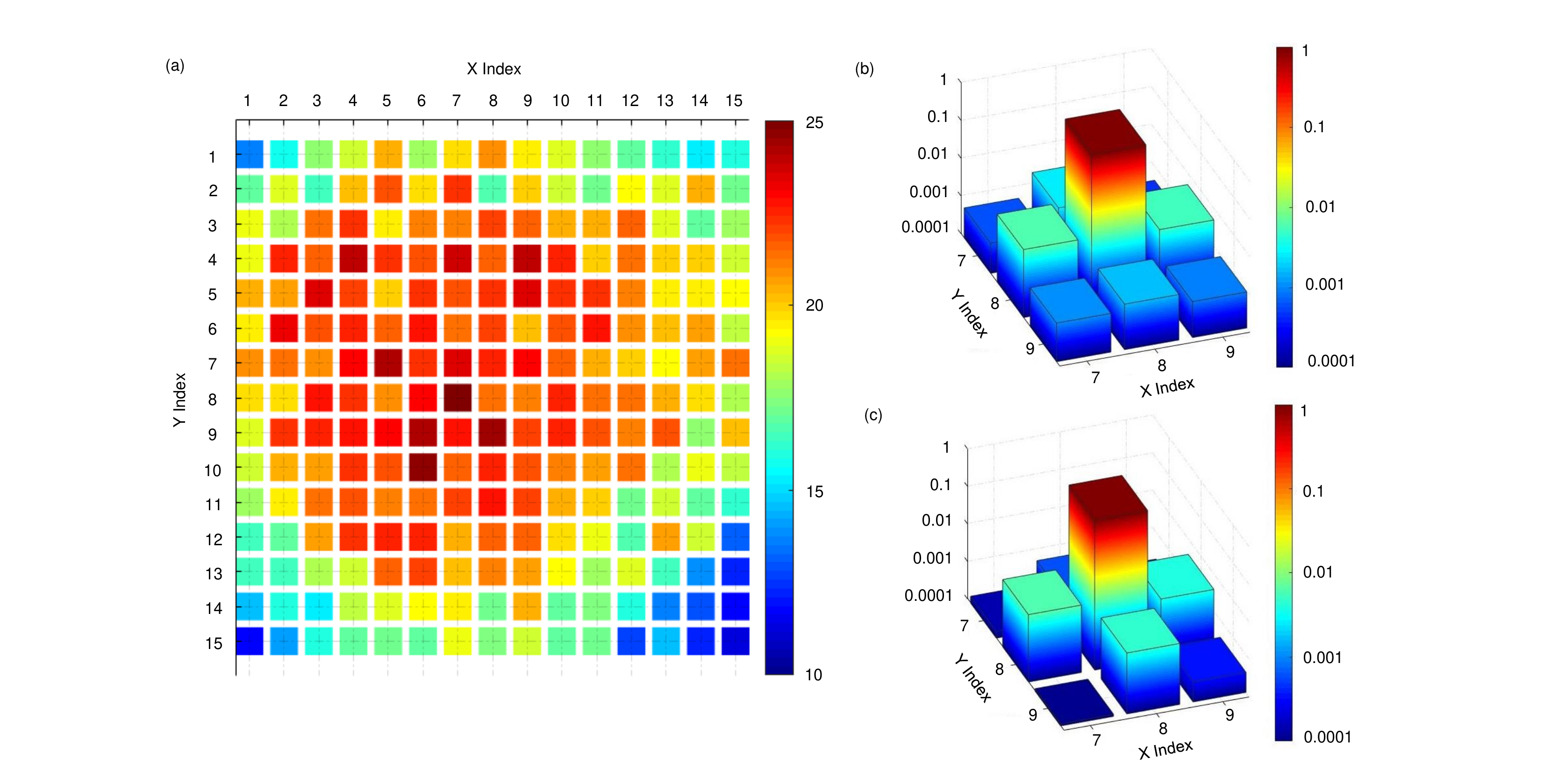}
\caption{\textbf{Quantum correlation and crosstalk errors in the $225$%
-cell quantum memory.} \textbf{a}, The measured intensity cross correlation $g_{\text{c}}$ between
the signal and the idler photons as a function of the cell index in $X$ and $%
Y$ directions for the 2D array of $15\times 15$ memory cells. The delay
between the read and the write pulses is $500$ ns. \textbf{b}, The measured
relative coincidence counts between the signal and the idler photons when
the write beam, the signal and the idler modes are addressed to one fixed
target cell (shown at the center of this figure with the cell index $\left(
8,8\right) $) while the read beam is scanned over the target and its
neighboring cells. The relative coincidence counts, with the target one
normalized to the unity, characterize the crosstalk errors of the read beams
on the neighboring cells. \textbf{c}, Similar to Fig. 2b, but with the write
beam scanned over the neighboring cells, which characterize the crosstalk
errors for the write beam.}
\end{figure}

\begin{figure}[ptb]
\includegraphics[width=17cm]{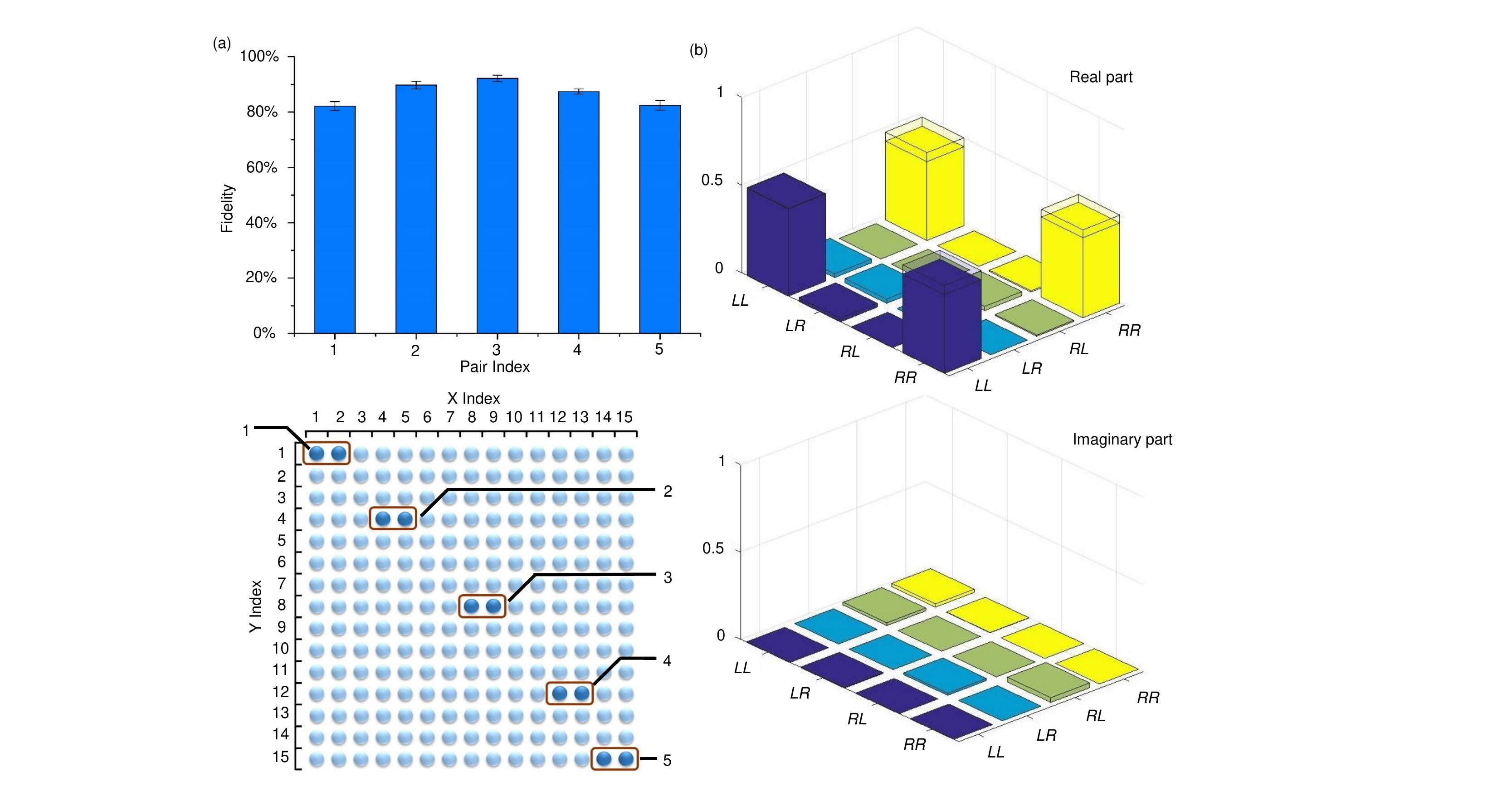}
\caption{\textbf{Quantum entanglement between the memory cells and the
signal photon.} \textbf{a}, The measured entanglement fidelity between the chosen pairs of
memory cells and the signal photon. The fidelity is calculated from the
experimental density matrix reconstructed through quantum state tomography
\textbf{\cite{27}}. The error bars correspond to one standard deviation
(s.d.). We determine the error bars by assuming a Poissonian distribution
for the photon counting statistics and propagate the error from the measured
quantity to the target property (fidelity here) through numerical Monte
Carlo simulation. The delay between the read and the write pulses is $500$
ns. The positions of the chosen pairs of cells in the 2D memory array is
shown in the lower part of the figure. \textbf{b}, The reconstructed density
matrix elements for the pair $3$ of the atomic cells with the signal photon.
The hollow caps denote the corresponding matrix elements for the ideal
maximally entangled state.}
\end{figure}

\subsection{Storage of atom-photon entanglement in memory cells}

To demonstrate storage of quantum entanglement in the memory cells, first we
generate entanglement between the memory cells and the outgoing signal
photons. For the DLCZ scheme, each write pulse generates with a small
probability $p$ a signal photon and an excitation in the corresponding
collective atomic mode \textbf{\cite{2}}. If the AODs equally split the
write beam into two paths L and R, when an excitation is generated, it is
equally distributed between the two paths, and the effective state in this
case is described by
\begin{equation}
\left\vert \Psi \right\rangle =\left( \left\vert \text{L}\right\rangle _{%
\text{s}}\left\vert \text{L}\right\rangle _{\text{a}}+e^{i\varphi
}\left\vert \text{R}\right\rangle _{\text{s}}\left\vert \text{R}%
\right\rangle _{\text{a}}\right) /\sqrt{2},
\end{equation}%
where $\left\vert \text{P}\right\rangle _{\text{s}}$ and $\left\vert \text{P}%
\right\rangle _{\text{a}}$ (P$=$L$,$R) denotes respectively a signal photon
and a collective atomic excitation in the path P, and $\varphi $ is a
relative phase which can be controlled by the electric signal applied on the
AODs. Through the AODs at the demultiplexing circuit, the states $\left\vert
\text{L}\right\rangle _{\text{s}}$ and $\left\vert \text{R}\right\rangle _{%
\text{s}}$ along the two paths can be combined into the output mode for
detection with an arbitrary weight function. We thus can detect the signal
photon in any basis $\cos \theta _{\text{s}}\left\vert \text{L}\right\rangle
_{\text{s}}+e^{i\phi _{\text{s}}}\sin \theta _{\text{s}}\left\vert \text{R}%
\right\rangle _{\text{s}}$ with the superposition weight function
characterized by the parameters $\theta _{\text{s}}$ and $\phi _{\text{s}}$.
Similarly, when we apply the read pulse on both paths L and R, the
collective atomic excitation $\left\vert \text{L}\right\rangle _{\text{a}}$
and $\left\vert \text{R}\right\rangle _{\text{a}}$, after conversion into
the corresponding idler photon states, are combined by the AODs at the other
side for detection in any basis $\cos \theta _{\text{a}}\left\vert \text{L}%
\right\rangle _{\text{a}}+e^{i\phi _{\text{a}}}\sin \theta _{\text{a}%
}\left\vert \text{R}\right\rangle _{\text{a}}$ with superposition parameters
$\theta _{\text{a}},\phi _{\text{a}}$. In our experiment, $\theta _{\text{s}%
} $ can be controlled by adjusting the amplitudes of the two frequency
components (corresponding to paths L and R respectively) in the
radio-frequency (RF) signal sent into the signal AODs, and $\phi _{\text{s}}$
can be controlled by adjusting the phase of the RF signal component
corresponding to the path R. The parameters $\theta _{\text{a}}$ and $\phi _{%
\text{a}}$ are controlled through the same method by adjusting the
amplitudes and phases of the RF signals sent into the idler AODs. By
choosing the appropriate parameters $\theta _{\text{s}},\phi _{\text{s}}$
and $\theta _{\text{a}},\phi _{\text{a}}$, we can perform quantum state
tomography of the underlying entangled state generated in experiments, and
the results are shown in Fig. 3 for different pairs of the memory cells from
the center of the ensemble to the edge. The entanglement fidelity $F_{\text{e%
}}$, defined as the overlap of the reconstructed experimental density
operator with a maximally entangled two-qubit state, is shown in Fig. 3a.
All the detected pairs give a high entanglement fidelity, far above the
criterion of $F_{\text{e}}>1/2$ for claim of entanglement.

To characterize the storage time of the quantum memory, in Fig. 4 we show
decay of the cross correlation $g_{\text{c}}$ and the entanglement fidelity $%
F_{\text{e}}$ for typically memory cells. The fits to the data for $g_{\text{%
c}}$ give a $e^{-1}$ decay time about $28$ $\mu $s. The entanglement
fidelity $F_{\text{e}}$ and the cross correlation $g_{\text{c}}$ are still
above the classical limit after $35$ $\mu $s storage time. The storage time
in current experiments is mainly limited by dephasing of the collective mode
caused by the motion of atoms and the small residual magnetic field gradient
when the magnetic-optical trap is shut off. By loading the atoms into an
optical lattice trap, it is possible to extend the storage time to the order
of seconds \textbf{\cite{21,22}}.

\begin{figure}[ptb]
\includegraphics[width=17cm]{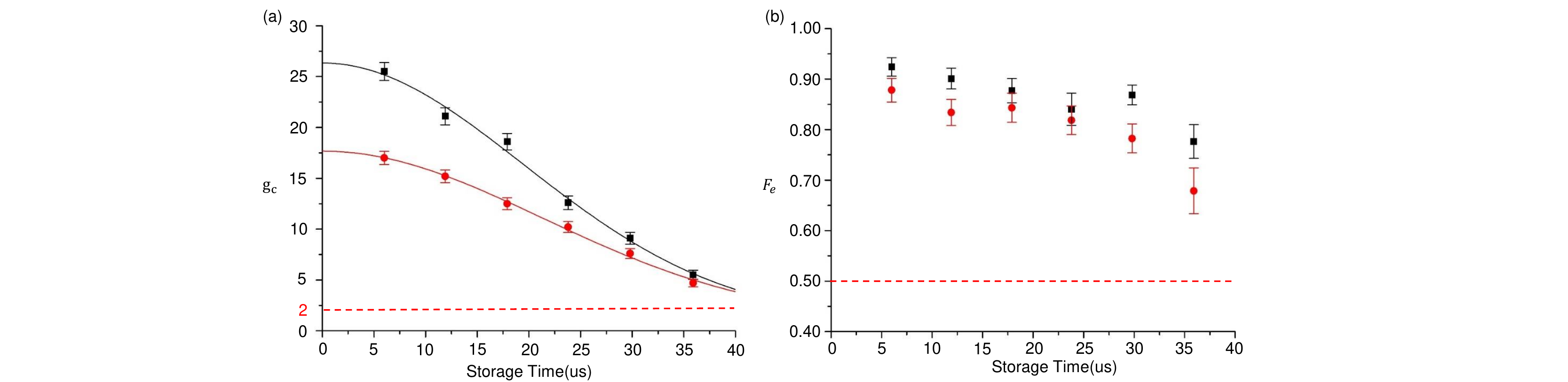}
\caption{\textbf{Measurement of quantum storage time in the multi-cell
quantum memory.} \textbf{a}, Decay of the cross correlation $g_{\text{c}}$ between the signal
and the idler photons with the storage time for a memory cell at the center
(squares) with the cell index $\left( 8,8\right) $ and at the edge (circles)
with the cell index $\left( 15,8\right) $. The solid lines correspond to
Gaussian fits $g_{\text{c}}=1+g_{0}\exp (-t^{2}/\tau ^{2})$ with $g_{0}=25.3$
($16.7$) and $\tau =27.5$ $\mu s$ ($30.1$ $\mu s$) for the upper (lower)
curves. The dashed line with $g_{\text{c}}=2$ corresponds to the limit above
which we have nonclassical correlation. \textbf{b}, Decay of the
entanglement fidelity $F_{\text{e}}$ with the storage time for entanglement
between a pair of memory cells at the center (squares) and the edge
(circles) of the 2D array and the flying signal photon. The dashed line with
$F_{\text{e}}=1/2$ corresponds to the limit above which we have genuine
quantum storage of entanglement impossible to be emulated by classical
operations. Error bars in (a,b) denote one s.d.}
\end{figure}

\section{Discussion}

Our experiment realizes a multiplexed quantum memory with hundreds of
individually accessible memory cells. The unprecedented large capacity of
the DLCZ quantum memory demonstrated in this experiment, together with the
programmable control of individual qubits in the memory cells with location
independent access time, opens up wide perspective for applications. This
type of memory is a critical device required for realization of multiplexed
quantum repeater networks for long-distance quantum communication \textbf{%
\cite{1,2,3,4,5}}. With many memory cells, individual and programmable
addressing, and efficient interface to the flying optical qubits, this
high-capacity quantum memory can also be used for demonstration of
many-particle entanglement \textbf{\cite{28,29}} and linear-optics quantum
information processing that requires memories \textbf{\cite{7,9}}, or as an
efficient node in the quantum internet \textbf{\cite{3}}.

\newpage

\section{Methods}

\subsection{Experimental setup}

We start by loading $^{87}$Rb atoms into a magneto-optical trap (MOT) inside
a vacuum glass cell. For cooling and trapping of the atoms in the MOT, we
use strong cooling beams red detuned to the D2 cycling transition $%
\left\vert 5S_{1/2},F=2\right\rangle \rightarrow \left\vert
5P_{3/2},F^{\prime }=3\right\rangle $ by $12$ MHz. The repumping laser,
resonant to the $\left\vert 5S_{1/2},F=1\right\rangle \rightarrow \left\vert
5P_{3/2},F^{\prime }=2\right\rangle $ transition, pumps back those atoms
which fall out of the cooling cycle. The diameter of the cloud in the MOT is
about $3.5$ mm and the temperature is about $300$ $\mu K$. We apply the
experimental sequence of the write and read pulses after we shut off the
MOT\ beams and the magnetic gradient coils and wait another $100$ $\mu$s.
For the data reported in Figs. 4 with longer storage time, we also include a
polarization gradient cooling (PGC) stage of $1$ ms before applying the
experimental sequence. The PGC is implemented by increasing the red detuning
of the cooling laser to $60$ MHz and reducing its intensity to half of the
value at the MOT loading stage. The repumping intensity is simultaneously
decreased to $0.005$ of the value at the loading phase and the magnetic
gradient coil is shut off. The temperature is decreased to about $30$ $\mu K$
by the PGC and the size of the MOT remains basically unchanged. After the
PGC some portion of the atoms are scattered to the $\left\vert
s\right\rangle \equiv \left\vert 5S_{1/2},F=1\right\rangle $ state, and a
repumping pulse of $100$ $\mu s$ long is applied after the PGC to pump all
the atoms back to the $\left\vert g\right\rangle \equiv \left\vert
5S_{1/2},F=2\right\rangle $ state. The ambient magnetic field is not
compensated during the experimental sequence, which induces an oscillation
in the retrieval efficiency of the collective atomic excitation by the
Larmor frequency \textbf{\cite{11,12,13,27a}}. In our experiment, the Larmor
period is $5.8$ $\mu s$ and the data in Figs. 4 are taken at integer periods
of the Larmor oscillation.

The experimental sequence begins with a write pulse of $100$ ns long, red
detuned by $10$ MHz to the D1 transition $\left\vert g\right\rangle
\rightarrow \left\vert e\right\rangle \equiv \left\vert 5P_{1/2},F^{\prime
}=2\right\rangle $ and focused to the atomic ensemble with a $60$ $\mu m$
Gaussian width. The signal photon is collected by a single mode fiber with a
focus Gaussian width of $35$ $\mu m$ at the atomic ensemble. If no signal
photon is detected, we apply a clean pulse of $100$ ns long and resonant to
the $\left\vert s\right\rangle \rightarrow \left\vert e\right\rangle $
transition, to pump the atoms back to the $\left\vert g\right\rangle $
state. The delay between each write pulse and clean pulse is $500$ ns, and
the write-clean sequence, with a total duration of $1$ $\mu s$, is repeated
until a signal photon is detected. When a signal photon is registered, we
stop the write-clean sequence. After a controllable storage time, a read
laser pulse, same as the clean pulse, is applied to retrieve the collective
atomic excitation to an idler photon mode. The conditional control of
write/read pulses is implemented by a home-made field-programmable gate
array (FPGA). The signal and the idler photons collected by the single-mode
fibers are detected by the single-photon counting module (SPCM), and the
events are registered by the coincidence circuit implemented with the FPGA.

\subsection{Multiplexing and de-multiplexing optical circuits}

Multiplexing and de-multiplexing optical circuits in our experiments are
achieved by placing four crossed acoustic-optical deflectors (AODs) in the
paths of the write/read beams and the signal/idler photon modes. The crossed
AODs, consisting of two AODs in orthogonal directions $X$ and $Y$, control
the beam deflection angle in these two directions. The crossed AODs,
together with the appropriately placed lens, achieve the addressing
configuration illustrated in Fig. 1. The AODs are placed on the focal points
of the two lens and the two lens are separated by two focal lengths with the
atomic ensemble located at the middle point between the two lens. The
directions of multiplexing and de-multiplexing AODs are adjusted so that the
frequency difference between the signal or idler photon along different
paths caused by the multiplexing AODs is canceled after combination by the
de-multiplexing AODs, which enables interference between different paths for
detection in superposition bases. To compensate the frequency difference
induced by different AOD driving frequencies for addressing different cells
of the memory array, a double-pass acoustic optical modulator (AOM) is
inserted in the write and the read laser beams to fix their detuning at the
atomic cells. As different optical paths in this multiplexing and
de-multiplexing circuits go through the same apparatus, the relative phase
differences between them are intrinsically stable and we do not need active
phase stabilization between different paths.

In our experiment, the mode-matching condition is fulfilled by
counter-propagation of the write and read beams and also the signal and
idler modes. To adjust matching of the spatial modes, we require the laser
beam emitted from the write (signal) fiber can be coupled into the read
(idler) fiber at the other side after transmission through all the optical
elements, no matter which atomic cells of the memory array they point to. In
our experiment, we achieve over $70\%$ coupling efficiency for all the $225$
optical paths addressing different memory cells.

The radio frequency (RF) signals to control the AODs are generated by two $4$%
-channel arbitrary-waveform generators (AWG). The four analog channels of
each AWG output RF with the same frequency and a controllable phase. The
four channels in one AWG control respectively the AODs for the write beam,
the read beam, the signal mode, and the idler mode, all addressed to the
same atomic memory cell. The RF with a different frequency generated by
another AWG then applies to these AODs to direct the beams to a different
memory cell. We combine the RFs of different frequencies from these two AWGs
into a single output for each of the four channels using a RF combiner.
These combined outputs are amplified to supply the driving field for the
four crossed AODs. The phase at each AOD can be controlled with a precession
about $0.1$%
${{}^o}$%
. To address different memory cells, the frequency of the RF for the crossed
AODs is scanned from $98.1$ MHz to $107.9$ MHz in step of $0.7$ MHz in both $%
X$ and $Y$ directions, which deflect the beams to $225$ different optical
paths pointing to the $15\times 15$ atomic memory cells.

To align the setup and fulfill the phase matching condition, we use the
following alignment procedure: 1. Set the RF frequency in all the $4$ pairs
of crossed AODs to aim at the middle cell of the 2D scanning array. 2. Send
a resonant probe laser ($|5S_{1/2},F=1\rangle \rightarrow
|5P_{1/2},F=2\rangle $) into the idler fiber and this beam
counter-propagates with the idler photon to be scattered by the atoms, then
steer the beam to achieve the highest optical depth. 3. Send another
resonant pumping beam ($|5S_{1/2},F=2\rangle \rightarrow
|5P_{1/2},F=2\rangle $) from the write fiber to overlap with the probe laser
aligned in step 2 to achieve the best electromagnetically induced
transparency (EIT) signal of the probe laser. 4. Couple the laser from the
idler fiber to the signal fiber, and couple the laser from the write fiber
to the read fiber. 5. Optimize the scanning ability by fine tuning of the
idler AODs, lens 1, atomic ensemble, lens 2 and the signal AODs to form a 4f
system (See Fig. 1 in the main text). Do the same tuning to the write AODs,
lens 1, atomic ensemble, lens 2 and the read AODs.

We use quantum state tomography to characterize the generated entanglement.
The quantum state tomography follows the standard procedure as described in
\textbf{\cite{27}}. We reconstruct the density matrix of the entangled state
of the atomic memory cells and the signal photon by the maximum likelihood
method, based on measurements in $4^{2}$ bases. The measurement bases we
used here are $|L\rangle _{\text{s}}$, $|R\rangle _{\text{s}}$, $(|L\rangle
_{\text{s}}+|R\rangle _{\text{s}})/\sqrt{2}$, $(|L\rangle _{\text{s}%
}-i|R\rangle _{\text{s}})/\sqrt{2}$ for the signal photon, and $|L\rangle _{%
\text{a}}$, $|R\rangle _{\text{a}}$, $(|L\rangle _{\text{a}}+|R\rangle _{%
\text{a}})/\sqrt{2}$, $(|L\rangle _{\text{a}}-i|R\rangle _{\text{a}})/\sqrt{2%
}$ for the atomic memory cells.

\section{Data availability}

The data that support the findings of this study are available from the
corresponding author upon request.

\textbf{Acknowledgements} We thank Y.-M. Liu, A. Kuzmich, L. Li, T. Tian,
and Y. Peng for helpful discussions. This work was supported by the
Tsinghua-QTEC Joint Lab on quantum networks and the National key Research
and Development program of China. LMD acknowledges in addition support from
the ARL CDQI program.

\textbf{Author Contributions} L.M.D. conceived the experiment and supervised
the project. Y.F.P., N.J., W.C., H.X.Y., C.L. carried out the experiment.
L.M.D. and Y.F.P. wrote the manuscript.

\textbf{Author Information} The authors declare no competing financial
interests. Correspondence and requests for materials should be addressed to
L.M.D. (lmduan@umich.edu).


\begin{thebibliography}{99}
\bibitem{1} Briegel, H. J., Dur, W., Cirac, J. I. \& Zoller, P. Quantum
repeaters: the role of imperfect local operations in quantum communication.
Phys. Rev. Lett. 81, 5932--5935 (1998).

\bibitem{2} Duan, L.-M., Lukin, M. D., Cirac, J. I. \& Zoller, P.
Long-distance quantum communication with atomic ensembles and linear optics.
Nature 414, 413--418 (2001).

\bibitem{3} Kimble, H. J. The quantum internet. Nature 453, 1023-1030 (2008).

\bibitem{4} Hammerer, K., Sorensen, A. S., Polzik E. S. Quantum interface
between light and atomic ensembles. Rev. Mod. Phys. 82, 1041-1093 (2010).

\bibitem{5} Sangouard, N., Simon, C., de Riedmatten, H., Gisin, N. Quantum
repeaters based on atomic ensembles and linear optics. Rev. Mod. Phys. 83,
33-80, 2011.

\bibitem{6} Collins, O. A., Jenkins, S. D., Kuzmich, A., Kennedy, T. A. B.
Multiplexed memory-insensitive quantum repeaters. Phys. Rev. Lett. 98,
060502 (2007).

\bibitem{7} Knill, E., Lafamme, R. \& Milburn, G. A scheme for efficient
quantum computation with linear optics. Nature 409, 46-52 (2001).


\bibitem{9} Barrett, S. D., Rohde, P. P. \& Stace, T. M. Scalable quantum
computing with atomic ensembles. New J. Phys. 12, 093032 (2010)

\bibitem{11} Chou, C. -W. \textit{et al.} Measurement-induced entanglement
for excitation stored in remote atomic ensembles. Nature 438, 828-832 (2005).

\bibitem{12} Chaneliere, T. \textit{et al.} Storage and retrieval of single
photons transmitted between remote quantum memories. Nature 438, 833--836
(2005).

\bibitem{13} Eisaman, M. \textit{et al.} Electromagnetically induced
transparency with tunable single-photon pulses. Nature 438, 837--841 (2005).

\bibitem{14} Julsgaard, B., Sherson, J., Cirac, J. I., Fiurasek, J., Polzik,
E. S. Experimental demonstration of quantum memory for light. Nature 432,
482--486 (2004).

\bibitem{15} Simon, J., Tanji, H., Ghosh, S., Vuletic, V. Single-photon bus
connecting spin-wave quantum memories. Nature Physics 3, 765-769 (2007).

\bibitem{16} de Riedmatten, H., Afzelius, M., Staudt, M. U., Simon, C.,
Gisin, N. A solid-state light matter interface at the single-photon level.
Nature 456, 773-777 (2008).

\bibitem{17} Lan, S.-Y. \textit{et al.} A Multiplexed Quantum Memory, Optics
Express 17, 13639-13645 (2009).

\bibitem{18} Lvovsky, A. I., Tittel, W. \& Sanders, B. C. Optical quantum
memory. Nature Photon. 3, 706-714 (2009).

\bibitem{19} Choi, K. S., Goban, A., Papp, S. B., van Enk, S. J., Kimble, H.
J. Entanglement of spin waves among four quantum memories. Nature 468,
412-416 (2010).

\bibitem{20} Saglamyurek, E. et al. Broadband waveguide quantum memory for
entangled photons. Nature 469, 513-518 (2011).

\bibitem{21} Dudin, Y. O., Li, L., Kuzmich, A. Light storage on the minute
scale. Phys. Rev. A 87, 031801(R) (2013).

\bibitem{22} Yang, S.-J., Wang, X.-J., Bao, X.-H., Pan, J. W. An efficient
quantum light--matter interface with sub-second lifetime. Nature Photonics
10, 381--384 (2016).

\bibitem{32} Dada, A. C., Leach, J., Buller, G. S., Padgett, M. J.,
Andersson, E. Experimental high-dimensional two-photon entanglement and
violations of generalized Bell inequalities. Nat Phys 7, 677-680 (2011).

\bibitem{33} Edgar, M. P. \textit{et al.} Imaging high-dimensional spatial
entanglement with a camera. Nat Commun 3, 984 (2012).

\bibitem{34} Ding, D.-S. \textit{et al.} High-dimensional entanglement
between distant atomic-ensemble memories. Light: Science and Applications 5,
e16157 (2016).

\bibitem{31} Hosseini, M. \textit{et al.} Coherent optical pulse sequencer
for quantum applications. Nature 461, 241-245 (2009).

\bibitem{30} Chrapkiewicz, R., Dabrowski, M., and Wasilewski, W.
High-Capacity Angularly Multiplexed Holographic Memory Operating at the
Single-Photon Level. Phys. Rev. Lett. 118, 063603 (2017).

\bibitem{23} Usmani, I., Afzelius, M., de Riedmatten, H., \& Gisin, N.
Mapping multiple photonic qubits into and out of a solid-state atomic
ensemble. Nature Commun. 1, 12 (2010).

\bibitem{24} Tang, J.-S. \textit{et al. } Storage of multiple single-photon
pulses emitted from a quantum dot in a solid-state quantum memory. Nature
Commun. 6, 8652 (2015).

\bibitem{25} Xia, T. Randomized benchmarking of single qubit gates in a 2D
array of neutral atom qubits. Phys. Rev Lett. 114, 100503 (2015).

\bibitem{26} Debnath, S., Linke, N. M., Figgatt, C., Landsman, K. A.,
Wright, K., Monroe, C. Demonstration of a small programmable quantum
computer with atomic qubits. Nature 536, 63-66 (2016).

\bibitem{27a} Kuzmich, A. \textit{et al.} Generation of nonclassical photon
pairs for scalable quantum communication with atomic ensembles. Nature 423,
731-734 (2003).

\bibitem{27} James, D.F.V \textit{et al.} Measurement of qubits. Rhys. Rev.
A. 64,052312 (2001).

\bibitem{28} Duan, L.-M. Entangling many atomic ensembles with laser
manipulation. Phys. Rev. Lett. 88, 170402 (2002).

\bibitem{29} Bodiya, T. P., Duan, L.-M. Scalable Generation of Graph-State
Entanglement through Realistic Linear Optics. Phys. Rev. Lett. 97, 143601
(2006).

\end{thebibliography}
\end{document}